\documentclass[conference]{IEEEtran}
\ifCLASSINFOpdf
\else
\fi
\usepackage[T1]{fontenc}
\usepackage[utf8]{inputenc}

\usepackage{amsmath,amssymb,mathtools}

\usepackage{booktabs,multirow,array,tabularx}

\usepackage{graphicx}
\usepackage{subcaption}

\usepackage[most]{tcolorbox}

\usepackage{cite}

\usepackage{xcolor}
\usepackage[hidelinks]{hyperref}
\usepackage{cleveref}

\usepackage{soul}
\sethlcolor{yellow} 
\usepackage{todonotes}

\usepackage{microtype}

\begin{document}
%
\title{When Stakeholder-centric Requirements
Engineering is Not Enough: An Action Research Study on Legacy System Modernisation}
%
%
%



\author{
Ruward S. Karper\IEEEauthorrefmark{1},
Damian A. Tamburri\IEEEauthorrefmark{1}\IEEEauthorrefmark{2},
Alessio Ferrari\IEEEauthorrefmark{3}\IEEEauthorrefmark{4},
Willem-Jan van den Heuvel\IEEEauthorrefmark{5}\\
\IEEEauthorrefmark{1}Jheronimus Academy of Data Science, DA 's-Hertogenbosch, the Netherlands\\
\IEEEauthorrefmark{2}TU Eindhoven, Eindhoven, the Netherlands\\
\IEEEauthorrefmark{3}University College Dublin, Dublin, Ireland \\
\IEEEauthorrefmark{4}CNR-ISTI, Pisa, Italy \\
\IEEEauthorrefmark{5}University of Tilburg, Tilburg, the Netherlands
}

%
%

\markboth{Journal of \LaTeX\ Class Files,~Vol.~14, No.~8, August~2015}%
{Shell \MakeLowercase{\textit{et al.}}: Bare Demo of IEEEtran.cls for IEEE Journals}
%



\maketitle

\begin{abstract}
    Legacy system modernisation is a major challenge in digital transformation, especially when organisations depend on long-lived, business-critical systems that are only partly understood. In such context, organisations must define future needs while determining what current systems actually do and which functions to retain, adapt, or replace. Modernisation is therefore not only a technical challenge but also a requirements engineering (RE) problem, shaped by stakeholder perspectives. This study examines how far stakeholder-centric RE can support gap analysis between the system-\textit{as-is} and the system-\textit{to-be} in a legacy modernisation context. We conducted an action research study in a multinational energy company engaged in system  modernisation. In the study, we applied stakeholder-centric RE practices, including stakeholder identification, semi-structured elicitation interviews, agreement-building through the Delphi method, and prioritisation with the MoSCoW method. The results show that this process was effective in producing requirements stakeholders generally viewed as understandable and correct, but less effective in achieving agreement on how elicited requirements mapped to legacy system functionality. The findings suggest that stakeholder-centric RE is necessary but not sufficient, pointing to the need for uncertainty-aware, iterative, and evidence-based modernisation practices that
combine stakeholder perspectives with manual and tool-assisted
analysis of legacy systems.
\end{abstract}


\section{Introduction}
\label{sec:intro}

A \emph{legacy system} is a business-critical system with high economic value that becomes increasingly difficult to maintain, owing to ageing technology, deteriorating documentation, and declining organisational expertise~\cite{Bisbal1997A}. 
According to Seacord et al.~\cite{seacord2003modernizing}, legacy system \textit{modernisation} concerns the evolution of a legacy system when conventional maintenance and enhancement can no longer achieve the desired system properties, and substantial renewal is required. Modernisation is a long standing challenge in software engineering, with solution proposal starting already in the nineties. Early work addressed this problem through reverse engineering, restoration, and step-wise replacement strategies~\cite{Visaggio1997Comprehending,Visaggio2001Ageing,Battaglia1998Renaissance,Hasselbring2004The}, while more recent work has focused on cloud migration and the decomposition of monolithic systems into microservices~\cite{grieger2016concept,marquez2015framework,fritzsch2018monolith,wolfart2021modernizing,mohottige2025reengineering}.
Industrial case studies focused on system re-implementation (e.g., changing programming languages  while keeping architecture and functionality~\cite{sneed2019reimplementing}), on cost assessment of modernisation decisions~\cite{martinez2017managing}, and on model-based support for reverse engineering.  Overall, most strategies proposed in the literature approach modernisation primarily as a \textit{technical} endeavour, usually motivated by the need to lower operational costs while improving interoperability and supporting better performance at scale~\cite{assunccao2025contemporary}. 

However, when considering the practitioner's viewpoint~\cite{khadka2014professionals}, the challenges experienced by professionals include a large portion of human-related issues. These include prioritisation difficulties, resistance from members of the organisation, as well as soft factors such as perspective misalignment between stakeholders. Some studies have therefore started to frame modernisation explicitly as a socio-technical process, in which knowledge and goals play a central role~\cite{Khadka2013Migrating}. This suggests that legacy system modernisation extends beyond architecture and refactoring to include a substantial requirements engineering (RE) dimension. Before introducing a technological transformation, organisations need to determine and agree on what the future system should do, what the current system already provides, and which existing functionality should be retained, adapted, or replaced~\cite{Khadka2013Migrating}. Despite the recognised relevance of RE for legacy system modernisation, limited studies exist that consider RE approaches to support this common industrial endevour. This paper aims to address this gap through an action research study conducted in a multinational company, where RE techniques have been applied to guide system  modernisation.  

Within RE, stakeholders are commonly approached as the primary repositories of domain and system information~\cite{NuseibehEasterbrook2000,Palomares2021Elicitation}. Accordingly, RE practice and research have long prioritized stakeholder-centric approaches, such as interviews, focus groups, and workshops~\cite{davis2006effectiveness}. 
In software modernisation projects, this emphasis naturally suggests an initial working assumption: if stakeholder knowledge and needs can be elicited and consolidated effectively, it should be possible to derive high-quality requirements for the system-\textit{to-be} and use them as a basis for comparing it with the system-\textit{as-is}. 
This paper examines that assumption in the context of a multinational energy company that must renew a legacy application used to analyze emissions data and produce statistics for sustainability analysis and external audits. The setting is characterised by multiple related systems deployed across different nations, which introduces heterogeneity and complicates the consolidation of system knowledge. In this context, we conducted an action research study~\cite{staron2020action} to assess whether stakeholder-centric RE practices can produce high-quality requirements for the target system and, in turn, whether these requirements provide a defensible basis for gap analysis against the capabilities of the original system, with the final goal of identifying reuse opportunities and missing functionality.


Methodologically, we select and apply a set of stakeholder-centric RE techniques, including semi-structured requirements elicitation inteviews, the Delphi  consensus-making technique\cite{Friedman2002Developing}, and the MoScoW method for requirements prioritisation~\cite{Suchetha2024}. The resulting requirements describing the system-\textit{to-be} were assessed as high quality by the stakeholders, in terms of understandability and correctness, i.e., relevance. However, when these requirements were subsequently used to determine which ones were actually implemented in the legacy system, substantial disagreement emerged among stakeholders, pointing to fragmented understanding of the existing system.

These findings suggest that, while stakeholder-centric RE, as applied in this study, can yield a high-quality specification of the intended system, the same artifacts may be insufficient to establish a shared, verifiable account of the baseline in complex legacy modernisation contexts. 
The observed phenomenon points to the limits of relying on stakeholder viewpoints alone and motivate the need to complement human- and process-centered RE with more explicit technical grounding. This primarily includes software analytics capabilities that can triangulate stakeholder accounts with evidence extracted from the system itself, since documentation is often scarce and much of the relevant knowledge remains embedded in the legacy system. 

Overall, while the legacy system literature focuses on technical aspects alone, and RE research mainly targets the human dimension, an integrated perspective is missing. 
RE researchers should work more closely with the mining software repositories and software architecture communities, to devise joint strategies to tackle the socio-technical nature of the legacy system modernisation problem. 

\smallskip
\noindent
Based on our findings and conclusions, the contributions of this paper are as follows:

\begin{itemize}
    \item Case-based evidence about the effectiveness of stakeholder-centric RE techniques in producing high-quality requirements in a legacy system modernisation context;
    \item Case-based evidence about the insufficiency of RE practices in producing requirements that can be effective to support strong conclusions on reuse potential;
    \item A set of lessons learned and reflections triggered by the experience.
\end{itemize}

\section{Related Work}
\label{sec:related}

Legacy system modernisation has long been recognised as a central problem in software engineering~\cite{assunccao2025contemporary}. The studies in the area can be roughly partitioned into the following categories: traditional studies focusing on reverse engineering and restoration~\cite{Visaggio1997Comprehending,Battaglia1998Renaissance,Hasselbring2004The}; works considering cloud migration, in particular through the decomposition of monolithic systems into microservices~\cite{fritzsch2018monolith,wolfart2021modernizing,mohottige2025reengineering}; industrial case studies~\cite{sneed2019reimplementing,martinez2017managing}; and surveys, including literaure reviews and interviews with practitioners~\cite{assunccao2025contemporary,khadka2014professionals}. 
In the following, we summarise relevant contributions in the different areas. 



\paragraph{Reverse engineering and restoration} Among early work on modernisation, Visaggio reported that comprehending and restoring an ageing system can simplify structure and improve maintainability, and later discussed symptoms and remedies for ageing in data-intensive systems \cite{Visaggio1997Comprehending,Visaggio2001Ageing}. Battaglia et al.\ introduced the ``renaissance method'', separating an initial decision of \emph{what to do} (renewal decisions) from \emph{how to do it} (evolving selected legacy features), and iterating between these concerns \cite{Battaglia1998Renaissance}. Later works focused on the transition problem, i.e., introducting strategies that enable coexistence and incremental replacement, as the legacy systems embed essential business logic and cannot be replaced abruptly. Among them, Hasselbring et al.\ propose the Dublo Pattern, which connects legacy business logic to a new tier to enable step-wise replacement while sustaining operations \cite{Hasselbring2004The}. While these early contributions testify to the long-standing nature of system modernisation, and provide foundational ideas and paradigms, they introduce limited process guidance for contemporary large-scale renewals.

\paragraph{Cloud Migration and Microservices} 
Cloud migration, also known as \textit{cloudification}, is the most common approach for system modernisation~\cite{assunccao2025contemporary}, and it has been addressed through a variety of frameworks in the literature~\cite{Hasan2023Legacy}.These frameworks mostly focus on the \textit{technical} restructuring needed to move legacy systems to the cloud. For example, Grieger et al.~\cite{grieger2016concept} start from conceptual modelling of the existing system and analyse migration patterns and strategies for individual components, while Marquez et al.~\cite{marquez2015framework}, with a focus on security, use automated tools to derive higher-level models and requirements from code and technical documentation, before selecting a cloud architecture. Still in the area of cloudification, one of the main trend is the decomposition of monolithic systems into microservices~\cite{assunccao2025contemporary}. Fritzsch et al.~\cite{fritzsch2018monolith} review refactoring approaches in the field, highlighting the context-dependent nature of the solutions, and providing a decision support scheme to select the most appropriate one based on the problem at hand. Wolfart et al.~\cite{wolfart2021modernizing} propose a roadmap for modernising legacy systems with microservices, outlining a generic process to guide cloudification, which starts from the analysis of the limitations of the legacy system, and arrives to the deployment and monitoring phase. Mohottige et al.~\cite{mohottige2025reengineering} provide a recent state-of-the-art review of methods, techniques and tools, for reengineering software systems into microservices, showing a strong emphasis on artifact-driven, static analysis solutions. Complementing these views, Carvalho et al.~\cite{carvalho2023re} report a survey with practitioners to identify the criteria used in practice to deal with modularity and feature variability when re-engineering legacy systems as microservices. 


\paragraph{Industrial Case Studies}
Given the relevance of the legacy system modernisation problem for practice, some of the works consider industrial case studies and experience reports. Sneed and Verhoef~\cite{sneed2019reimplementing} supports \textit{re-implementation} as a practical strategy between direct conversion and complete redevelopment, preserving the functional architecture of the legacy system while replacing its technical implementation (e.g., replacing COBOL with Java). The paper describes two re-implementation case studies, observing that, while code quality improved, complexity of the code also increased. In a financial-services case study,  Crotty and Horrocks~\cite{martinez2017managing} develops a meta-assessment model to analyse the costs of maintaining legacy systems and to support managerial decision making about modernisation. The authors apply the approach to FinCo, a large UK financial services company, receiving confirmation of the utility of the method. Garc{\'\i}a-Borgo{\~n}{\'o}n et al.~\cite{garcia2023lessons} present lessons learned from applying model-based reverse engineering to large industrial legacy systems, focusing on how legacy knowledge can be reconstructed from existing artefacts before further modernisation steps are planned. They conclude that the applied method brings an average saving of 75\% of the eﬀort needed
to understand how a legacy system is built and technological debt analysis, and 99\% cost improvement in terms of producing documentation, compared to traditional approaches.



\paragraph{Surveys} 

Assun{\c{c}}{\~a}o et al.~\cite{assunccao2025contemporary} provide a recent literature review on system modernisation, identifying drivers and challenges. The paper highlights that most of the published works focus on reducing operational costs, and that tool support is the main challenge recognised by the studies. Interestingly, challenges associated with RE and stakeholders are not prominent in the surveyed literature. On the other hand, the socio-technical dimension appears to be slightly more present in the multi-vocal literature review by Hogan et al.~\cite{olanrewaju2024investigating}, which remarks that one of the main challenges resides in integrating the novel system with the existing process, which includes tasks and stakeholders. The broader socio-technical framing is instead strongly present in the survey with practitioners by Khadka  al.~\cite{khadka2014professionals}. This highlights that the main challenges of system modernisation include the fragmented expertise and lack of knowledge, the difficulty in understanding the business logic and rationale of the legacy system, and ``soft'' modernisation factors, including communication issues between the management and the technical teams, thus posing stakeholders at the center of the modernisation problem. A concrete example of this perspective is provided by the banking case study of Khadka et al.~\cite{Khadka2013Migrating}, which places explicit emphasis on governance and stakeholder involvement throughout the migration process. Their approach begins with the establishment of a migration committee comprising both technical and managerial stakeholders, followed by the definition of a target architecture grounded in business goals, a gap-analysis phase, and the subsequent realisation of the migration. In this way, Khadka et al.\ make explicit that modernisation is not merely an architectural transformation problem, but a socio-technical endeavour in which technical decisions are closely intertwined with stakeholder participation and the continuous elicitation of organisational knowledge.

\paragraph{Research Gap} Most of the previous work treat  modernisation mainly as a system architecture or refactoring problem, typically neglecting the RE dimension in general and the relevance of stakeholders in particular. 
The only exception is the case study by Khadka et al.~\cite{Khadka2013Migrating}, where stakeholders have a prominent role. However, the paper mainly focuses on lessons learned, failing to provide detailed reproducible steps, and, although the approach is in line with the RE spirit (e.g., involve stakeholders, goals before solutions, plan before acting), it does not appear to follow structured RE practices. 
Overall, despite the relevance of the RE dimension in legacy system modernisation, studies that address this problem explicitly from an RE perspective remain scarce.  

\paragraph{Contribution} Compared to previous work, our contribution investigates whether following stakeholder-centric RE approaches from the literature can facilitate the process of legacy system modernisation. In addition, this study shows that, while the RE practices as applied in this modernisation process produced high-quality requirements, they were not sufficient to support a reliable comparison between the legacy system and the target system.

\section{Research Design}
\label{sec:method}

The overarching goal of this study is to understand the extent to which, established stakeholder-centric RE practices, can support legacy system modernisation in a large (5000+ employees) multinational company active in the energy and power services sector. To address this goal, we followed the action research steps described by Staron~\cite{staron2020action}. Accordingly, the study encompasses a constant presence of industrial stakeholders embedded in the phases below:

\begin{itemize}
    \item \textbf{Diagnosis:} the goal of this phase is to understand the problem context and identify the underlying challenges. In our case, $<$anonymous company$>$ contacted the authors with the aim of modernizing their reporting system. To capture and scaffold the problem, informal meetings were held with company representatives in the first weeks of collaboration. The problem unfolded as a stakeholder-centered RE problem, where (i) stakeholder identification, requirements elicitation as well as specification were required to understand the requirements for the novel system, and (ii) gap analysis\footnote{In this study, gap analysis refers to the assessment of the extent to which the requirements identified for the target system were already supported by the legacy system. Specifically, we operationalised gap analysis as a comparison between the requirements of the system-\textit{to-be} and the features of the system-\textit{as-is}, aimed at identifying which requirements were already fully, partially, or not supported by the legacy system. More broadly, the term is also used throughout the text to denote any general comparison between the two system versions.} was needed to understand which features of the legacy system could be reused to satisfy those requirements, and which features were missing.   
    \item \textbf{Action planning:} this phase consists in formalising the research questions (RQs) that should be addressed to solve the diagnosed problem, and defining the actual treatment to be applied to address the RQs. Furthemore, the phase also defines the data collection and analysis procedure to be carried out during the subsequent action taking and evaluation phases (cf. below). In our case, it consisted of: (i) selecting stakeholder-centric RE practices oriented to produce requirements specifications that could be reliably used to perform gap analysis; (ii) formulating RQs to evaluate the effectiveness of the selected RE practices for the problem at hand; 
    (iii) defining a procedure to apply those practices; (iv) identify metrics to evaluate whether the application of the RE practices was successful, together with procedures to collect data for the metrics.  
    \item \textbf{Action taking:} in this phase, we applied the different RE practices selected during the action planning phase in a six-month time span involving relevant stakeholders of the company. The result of this activity was a requirement specification that was used in the evaluation phase. The different practices required a minimum of 2-3 days up to 2-3 weeks, followed typically by 2-3 days of synthesis in the form of a progress presentation/report by the first author to the other co-authors.
    
    \item \textbf{Evaluation:} this phase consists of providing answers for the RQs. 
    The specified requirements were evaluated by examining how well they supported modernisation decisions. To this end, we involved stakeholders in a questionnaire to (i) evaluate the quality of the resulting requirements, and (ii) perform a gap analysis using those requirements. 
    
    \item \textbf{Specifying learning:} after the experience, the authors and stakeholders discussed the lessons learned in a plenary workshop. 
\end{itemize}




\section{Diagnosis: Modernisation as a RE Problem}
The study was conducted at $<$anonymous company$>$, a multinational energy company operating in Sweden, Denmark, Germany, the Netherlands, and the United Kingdom. The work took place in $<$anonymous company$>$'s Heat department, which is responsible for heat generation and related assets, such as heat generation plants across Sweden, Germany, and the Netherlands. Within this department, the researchers collaborated closely with the Asset Information Management (AIM) team, which manages the digital information associated with these assets.

As an energy company, $<$anonymous company$>$ must not only generate sufficient power and heat for its customers, but do so in a sustainable manner, with minimal emissions and reduced dependence on fossil fuels. At the time of writing, $<$anonymous company$>$'s stated mission was to become fossil-fuel-free within one generation. This implies an obligation to report fuel and energy consumption, as well as heat and power production, to internal and external stakeholders, including plant operators and governmental actors. These reports support both operational improvement and regulatory compliance. For this purpose, the AIM team produces technical reports on the consumption and generation performance of $<$anonymous company$>$'s heat assets.

These reports depend on technical time-series data, for example emissions data recorded per hour or per day. The reliability and quality of this data are critical, since the reports must provide a faithful representation of the performance of $<$anonymous company$>$'s assets. In addition, external audits require the underlying reporting process to be transparent and accountable.

To ensure the quality of the data used for reporting, different countries within the company relied on different legacy applications. These applications validated data by detecting and replacing missing values, checking plausibility, and performing calculations needed to derive reportable quantities from raw asset data. For example, CO\textsubscript{2} emissions could be calculated by combining time-series data for natural gas with its emission factor and calorific value.

In recent years, $<$anonymous company$>$ has pursued a digital strategy that first focused on moving data to the cloud to improve accessibility, maintainability, and management, and then on renewing the legacy applications used for reporting. The first phase had already been completed when this study began: the relevant reporting data had been migrated to the infrastructure of a chosen cloud vendor.

The company had therefore entered the second phase of its strategy, namely the renewal of the legacy reporting applications and, potentially, their migration to the cloud. Across the countries involved, the existing applications were regarded as end-of-life. They had been developed years earlier, and both users and management considered them increasingly inadequate for current business needs. Because the reporting process was business-critical, the company wished not only to renew these systems but also, where possible, to harmonise them across countries.

However, the organisation faced difficulties in initiating the renewal effort effectively. Although there was consensus that the systems should be renewed, it remained unclear what the target situation should be and to what extent it differed from the functionality already offered by the legacy systems. In particular, the company lacked a systematic basis for deciding which legacy functionality should be retained, which should be changed, and which new functionality should be introduced.

\textit{Problem Formulation.} From an RE perspective, this was a legacy system modernisation problem in which the organisation needed to determine which legacy functionality should be retained, adapted, or replaced, but lacked a sufficiently explicit and stakeholder-validated specification of the target system against which the legacy applications could be assessed. 
We therefore formulated the problem as an RE problem concerned with (i) eliciting and formalising stakeholders' needs for the target system and (ii) using the resulting requirements specification to support gap analysis between the system-\textit{as-is} and the system-\textit{to-be}.

\section{Action Planning}
\label{sec:planning}

Given the diagnosed problem, the intervention focused on designing an RE process capable of producing a requirements specification that could serve as a reliable basis for system modernisation. We adopted a stakeholder-centric approach because legacy renewal initiatives carry a substantial risk of reproducing existing system functionality without sufficiently examining whether it still reflects present stakeholder needs. By centring the process on stakeholder goals and knowledge rather than on the structure of the existing system, the intervention aimed to reduce the risk of uncritical legacy carryover, a problem defined by Alexandrova et al. as the \textit{legacy problem}~\cite{Alexandrova2015The}. This was particularly important in the present case, where the organisation sought not only renewal but also harmonisation across multiple countries and organisational contexts. Based on this rationale, we defined the following overarching goal.

\textbf{Overarching goal.} \textit{To investigate to what extent stakeholder-centric RE practices can produce a requirements specification that supports gap analysis between a legacy system (system-as-is) and a target system (system-to-be) in a multinational system modernisation context.}

To address this goal, we selected a sequence of stakeholder-centric RE practices aimed at progressively moving from stakeholder knowledge to a structured requirements specification. These techniques include: (i) stakeholder identification according to Hujainah et al.~\cite{Hujainah2018Software} and Pacheco and Garcia~\cite{Pacheco2012A}, (ii) semi-structured requirements elicitation interviews, based on general guidelines~\cite{Alshenqeeti2014Interviewing,Barbour2005Interviewing} and RE-specific ones~\cite{donati2017common,ferrari2019learning}, and thematic analysis of the interview transcripts based on Braun and Clarke~\cite{Braun2012Thematic} to identify preliminary requirements; (iii) a consensus workshop based on the Delphi method~\cite{Friedman2002Developing}; (iv) requirements specification through user stories, widely used in RE research and practice; and (v) requirements prioritisation based on the MoSCoW method~\cite{Suchetha2024}. The specific practices were opportunistically selected based on the experience of the research team. Given the time constraints of the action research study, identifying the most suitable practices, learning them, and applying them was not considered feasible. Although this has a substantial impact on the internal validity of the findings, these shortcomings are inherently part of any field experiment~\cite{stol2018abc}\footnote{The ABC for software engineering research by Stol et al.~\cite{stol2018abc} classifies action research studies as field experiments.}.  

The sequence of practices was inspired by the work of Khadka et al.~\cite{Khadka2013Migrating}, which starts from the creation of a migration committee, elicits needs and architecture for the novel system, and then performs gap analysis. However, it not intended as a prescriptive RE lifecycle, but as a study-specific operationalisation of established RE activities for the purposes of legacy system renewal. In line with standard RE references~\cite{pohl2025requirements}, which describe the RE process as progressing from elicitation and analysis toward specification and validation, we organised the intervention so as to move from stakeholder identification and need elicitation to agreement, specification, and prioritisation. This ordering was considered appropriate in our context because it enabled a gradual transition from dispersed stakeholder knowledge to a structured requirements specification that could subsequently support a comparison between the envisioned system and the legacy system. The concrete application of the selected  practices in the industrial setting is described in the action taking phase.

To evaluate the extent to which this intervention achieved its goal, we formulated the following research questions, aiming to assess the quality of the produced requirements, and their usefulness to identify reuse opportunities and missing functionality.

\textbf{RQ1 (Requirements quality).}
\textit{To what extent do stakeholder-centric RE practices lead to a requirements specification that stakeholders assess as understandable and correct?}

\textbf{RQ2 (Gap-analysis consistency).}
\textit{To what extent does the resulting requirements specification enable consistent stakeholder judgements about whether each requirement is already supported by the legacy system?}

The evaluation strategy was also defined during action planning. RQ1 was operationalised through stakeholder assessment of requirement understandability and correctness, while RQ2 was operationalised through stakeholder classification of whether each requirement was already supported by the legacy system. In the following, we report the details of the data collection and analysis procedures.

\paragraph{RQ1---Data Collection} RQ1 focuses on understandability and correctness of the requirements. According to Davis et al.~\cite{Davis1993Identifying}, a requirement specification is understandable ``if all classes of SRS readers can easily comprehend the meaning of all requirements'', and it is correct if ``every requirement represents something required of the system to be built, i.e., every requirement [...] contributes
to the satisfaction of some need.''. Understandability and correctness form a subset of the quality criteria for a requirements specification from Davis et al.~\cite{Davis1993Identifying}, which were selected as they were considered by the company as the most relevant quality dimensions for the renewal problem at hand. In particular, the company needed a requirements specification that stakeholders could understand consistently and recognise as reflecting actual needs. Other quality dimensions, e.g., modifiability~\cite{Davis1993Identifying}, were considered less directly relevant to this specific objective.

To collect the evaluation data, an online questionnaire was designed and shared with the involved stakeholders, including representatives of users from the different countries, management, and data analytics. For each requirement in the specification, stakeholders were asked whether they understood the requirement (\textit{``Do you understand what this requirement means?''}), and whether they considered it to address a need (\textit{``Does this requirement satisfy a need from
your perspective?''}). Understandability and correctness were assessed through a binary judgement, and, in the case of correctness, respondents had the possibility to indicate that a requirement was not applicable to them. This was not allowed when assessing understandability, since it was considered important that all requirements are understandable also to stakeholders to whom the requirement is not directly applicable. Since the requirements were going to be expressed as user stories, a standard high-level format specifically oriented to non-techincal stakeholders, this restriction was considered reasonable. It should be highlighted that simply asking whether a requirements was understandable, does not imply that the respondent actually understood it correctly, as phenomena of incorrect subconscious disambiguation could have occurred~\cite{ferrari2016ambiguity}. However, given the limited time of the stakeholders, asking to rephrase, or introducing other strategies for checking actual understanding, was not considered an option, as it would limit the number of responses. Furthermore, although binary assessment does not account for the shaded nature of the constructs considered, this was preferred to a Likert scale as it forces a clear-cut decision, which are preferred in a time-constrained context. 

\paragraph{RQ2---Data Collection} Once requirement quality had been assessed, the same questionnaire was used to support the gap-analysis task. For each requirement, stakeholders were additionally asked whether the requirement was already covered by the current legacy systems (\textit{``Is this requirement already covered by the functionality of the current systems?''}). Possible answers distinguished between full coverage, no coverage, and partial coverage; when stakeholders indicated coverage or partial coverage, they could also specify the relevant legacy system or explain the reason for partial support. This enabled the collection of a dataset that captured, for each requirement, stakeholders' views on legacy system coverage.

\paragraph{Data Analysis} In both steps, the consistency of stakeholder judgements was analysed through inter-rater agreement using Krippendorff's Alpha~\cite{Krippendorff2011Computing}. In addition to agreement, the evaluation also considered descriptive outcome measures: for RQ1, the proportion of requirements judged understandable and correct, respectively; and for RQ2, the proportion of requirements judged fully or partially supported by the legacy systems. The combination of requirement-quality results and functionality-mapping agreement further enabled us to interpret whether ambiguity in the results primarily stemmed from the requirements specification itself, from differing understandings of legacy-system functionality, or from both.

\section{Action Taking}
\label{sec:actiontaking}

The selected RE practices were enacted in sequence, starting from identification of stakeholders until requirements specification and prioritisation. In the following, we report details on the implementation of the planned process.

\subsection{Identification of Stakeholders}

Stakeholder identification was carried out by combining the stakeholder categories proposed by Hujainah et al.~\cite{Hujainah2018Software} with the practices of Pacheco and Garcia~\cite{Pacheco2012A} for stakeholder identification. We first examined the organisational structure of the AIM team and identified candidate stakeholders based on their current responsibilities in relation to the validation and calculation system. These candidates were then assigned to three categories: \emph{functional-beneficiary}, \emph{technical}, and \emph{commercial}. Functional-beneficiary stakeholders were identified among the main users and domain experts of the existing systems; technical stakeholders were identified among the actors responsible for data management, analytics, and technical solution design; and commercial stakeholders were identified among the relevant management roles. This initial classification was subsequently discussed with AIM management, which led to the inclusion of additional technical actors outside AIM, namely the solution designer, solution architect, legacy system manager, and a developer representative, as these roles were considered necessary to represent knowledge about the design, operation, and historical development of the legacy systems. The final list of 14 identified stakeholders is shown in Table~\ref{tab:stakeholders}.

\begin{table}[h]
\centering
\caption{Identified stakeholders}
\label{tab:stakeholders}
\begin{tabular}{ll}
\hline
\textbf{Stakeholder group} & \textbf{Stakeholder} \\
\hline
Functional-beneficiary & Technical reporters NL \\
 & Specialist reporting NL \\
 & Technical reporters GER \\
 & AIM SWE \\
\hline
Technical & Data management \\
 & Data analytics \\
 & Specialist data NL \\
 & Solution designer \\
 & Solution architect \\
 & Legacy system manager \\
 & Developer representative \\
\hline
Commercial & Manager AIM \\
 & Manager AIM-NL \\
 & Manager AIM-GER \\
\hline
\end{tabular}
\end{table}

The functional-beneficiary group comprised the main users of the legacy systems and roles with reporting-related domain knowledge in the Netherlands, Germany, and Sweden. The technical group comprised roles responsible for data management and analytics within AIM, complemented by actors with design, development, and operational knowledge of the legacy systems. The commercial group consisted of the relevant management roles, which represented organisational goals, priorities, and external business expectations.

\subsection{Requirements Elicitation Interviews and Workshop}

Requirements elicitation was conducted through semi-structured interviews with the stakeholders listed in Table~\ref{tab:stakeholders}. Semi-structured interviews were selected because they provide enough structure to ensure coverage of the main topics while leaving room for stakeholders to elaborate on their work practices, needs, and concerns. The interviews were designed following established guidance on qualitative interviewing~\cite{Alshenqeeti2014Interviewing,Barbour2005Interviewing}, also considering RE-specific guidance to avoid common pitfalls~\cite{donati2017common,ferrari2019learning}. The interviews were conducted in a conversational style, with the aim of creating a comfortable setting in which participants could speak freely and the interviewer could focus on listening and probing for clarification when needed. To reduce power imbalance and anchor the discussion in the participants' expertise, each interview started with questions about the interviewee's role, daily work, and involvement with the existing system, before discussing the legacy system's problems and the expectations from the modernised system.

Interview guides were tailored to the different stakeholder groups and consisted of broad, open-ended questions. While all interviews mainly addressed the overall goal of understanding the expectations for the renewal effort as well as the characteristics of the current system, the emphasis varied by role. Users were asked primarily about their current use of the system (e.g., \textit{``How do you use the system in your daily work?''}), difficulties encountered (e.g., \textit{``What is the first thing you would change about the
system?''}), and desired functionality (e.g., `\textit{`What do you miss in the system that you consider
essential?''}); management stakeholders were asked about business goals (\textit{`` What is from a management perspective the most important goal of roadmap 2.0?''}), priorities (\textit{``What would be the most important factors when choosing between multiple possible new applications?''}), and expected outcomes of the renewal (\textit{``What value will roadmap 2.0 have within the company
as a whole?''}); and technical stakeholders were asked about data flow (\textit{``What is your vision of what the data flow would ideally
look like?''}), and other data-related aspects, using a more technical language (\textit{``The users already indicate a number of important matters: transparency, speed, user friendliness, reliability,
flexibility. If you hear these topics, how do you think
we can best deal with them from a data perspective?
Think about data accessibility, data transparency, single
point of truth, etc.''}). The objective of the interviews was to elicit an initial understanding of the stakeholder's knowledge on the current system, as well as their needs, so that these could be translated into preliminary requirements. 

The interview data were analysed using thematic analysis following Braun and Clarke~\cite{Braun2012Thematic}. Interview answers were first coded by summarising or aggregating relevant statements into short descriptive codes. These codes were then clustered into \textit{themes} representing recurring needs, concerns, and expectations regarding the future system, and that took the form of informal preliminary requirements (e.g., ``track changes in system'', ``statistics on data and system load'', ``collection of manual inputs from plant technician''). As the analysis progressed, the themes were further organised into functionality groups that captured the main areas of the system under discussion, namely: data ingestion, validation, calculation, export, plant characteristics, data analytics, and general. These groups helped structure the problem space and clarify the system's intended scope. The process produced a total of 40 themes within the seven functionality groups.

The preliminary findings from the thematic analysis were then presented to the \textit{full} stakeholder group, i.e., 14 participants, in a workshop session, carried out based on the Delphi method~\cite{Friedman2002Developing} for consensus making. Specifically, for each theme, each stakeholder was engaged in a SWOT analysis, in which they were asked to assess strenghts (in terms of potential relevance and impact), weaknesses (how much the theme was ill-defined), opportunities (in terms of added value for the business), and threats (in terms of business risks). Three rounds of discussion were performed until consensus was reached. While during the session the themes were not formalised as user stories, potentially hampering their interpretability, the interactive nature of the session facilitated shared understanding. This session served to validate the emerging interpretations, harmonise understanding across stakeholders, and refine the next step of the planned process. 


\subsection{Requirements Specification}

Based on this discussion, the themes were converted into an a list of 105 requirements and circulated for validation and prioritisation. The requirements were formalised through user stories, using the format \emph{As a \textless Role\textgreater{} I want to \textless do Action\textgreater{} because \textless Goal\textgreater{} by \textless How\textgreater{}}. Example requirements are: ``As a reporter I want to have normalized data available for comparison of data in equal units and granularity by converting data into the same dimensions’’; and ``As a reporter I want to be notified when there is no connection with PI to estimate the impact of connectivity issues on technical reporting by receiving an active notification when connectivity issues occur’’. The final dataset contained 16 entries/topics---or \emph{epics} in the Agile parlance \cite{Schuh2018Agile}---covering 105 requirements.

\subsection{Requirements Prioritisation}

All stakeholders were invited to take part to the prioritisation process. Requirements were prioritised using the MoSCoW method, selected for its simplicity and its suitability for use in the early stages of a project. Each stakeholder was asked to rate every requirement as \emph{Must}, \emph{Should}, \emph{Could}, or \emph{Won't}, based on their view of the system. This produced a dataset containing a distribution of prioritisation scores for each requirement, which could then be used to determine which aspects of the system should be elaborated first. The idea was to understand the requirements that were expected to be implemented in a minimum viable product (MVP) of the envisioned system, i.e., the first basic version of the target system that was expected to be completed enough to demonstrate the value it would bring to the company~\cite{Moogk2012Minimum}. 

Although the list of requirements was distributed to all stakeholders, in practice it appeared most suitable for management, end-users, and the data department, which amounted in total to 9 people. The IT developers and architects indicated that they did not consider themselves the appropriate actors to rate the requirements, as their role was to design and build the solution rather than to express preferences about what the system should do.

To enable a fair comparison across perspectives, stakeholder input was weighted at group level. For example, the German users were represented by a single participant, who had collected input from the other German users and reflected this in his own ratings, whereas the Dutch users were represented by four participants. To balance this difference in representation, the German user's responses were counted four times. Although this gave that participant greater influence, this adjustment was not communicated in advance and therefore could not have influenced the ratings themselves; moreover, it was considered the most practical way to ensure equal representation of the two user groups. Similarly, management was given the same overall weight as the combined user group, as it was considered more appropriate to balance stakeholder groups than individual stakeholders in order to reflect the different perspectives involved.

Not all invited stakeholders submitted their ratings. The Swedish user base did not complete the prioritisation, and only two of the three management representatives responded. These stakeholders had been informed that, if they did not respond by the deadline, it would be assumed that their views were sufficiently represented by their colleagues' input. On that basis, the collected ratings were considered an acceptable representation of the stakeholder perspectives involved. An estimate response rate was achieved around 58\%, which is well above typically acceptable levels.



Agreement on the priority was rather low, as for less than 20\% of the requirements at least 80\% of the stakeholders actually agreed on the precise rating. Given the large number and variety of stakeholders this should have been, however, expected. For this reason it was decided to discard any requirements that had less than half of the votes include ``must’’ or ``should’’. The final list of retained requirements for the MVP consisted of 56 items.

\section{Evaluation}
\label{sec:results}

This section reports the results of the evaluation and answers the two research questions defined in the action planning phase. 

\subsection{RQ1: Requirements Quality}

RQ1 asked to what extent the stakeholder-centric RE practices led to a requirements specification that stakeholders assessed as understandable and correct. To answer this question, the requirements were evaluated through the stakeholder questionnaire introduced in Sect.~\ref{sec:planning}, and inter-rater agreement was analysed using Krippendorff's Alpha. Table~\ref{tab:alpha-values} reports the agreement values obtained for the three selected quality dimensions, together with the value for functionality mapping used later to answer RQ2.

\begin{table}[ht]
\centering
\caption{Agreement values}
\label{tab:alpha-values}
\begin{tabular}{lc}
\hline
\textbf{Agreement on} & \textbf{Alpha value} \\
\hline
Understandability & 0.94 \\
Correctness & 0.90 \\
Functionality mapping & 0.45 \\
\hline
\end{tabular}
\end{table}

The results show very high agreement on understandability and correctness, with alpha values of 0.94 and 0.90, respectively, reflecting the effectiveness of the different viewpoint sharing workshops.
Agreement alone, however, does not indicate whether the requirements were judged positively or negatively. We therefore also examined the extent to which the requirements satisfied the selected quality dimensions. Following Davis et al.~\cite{Davis1993Identifying}, quality was interpreted as the proportion of requirements satisfying each measure. Because categorical ratings do not allow meaningful averaging, we report two interpretations: a \emph{tight} interpretation, in which all raters had to agree, and a \emph{lenient} interpretation, in which a majority judgement was sufficient.

\begin{table}[ht]
\centering
\caption{Requirement quality and functionality mapping}
\label{tab:srs-values}
\begin{tabular}{lccc}
\hline
\textbf{Measure} & \textbf{Rating ratio} & \textbf{Tight} & \textbf{Lenient} \\
\hline
Understandability & 30.5--1 & 80\% & 100\% \\
Correctness & 18.4--1 & 66\% & 100\% \\
Functionality mapping & 1.34--1, 0.59--1 & 0\% & 50\% \\
\hline
\end{tabular}
\end{table}

Table~\ref{tab:srs-values} shows that the resulting specification performed strongly on understandability and, to a sufficient extent, also correctness. Under the lenient interpretation, all requirements were judged understandable and correct by a majority of stakeholders; under the tight interpretation, the proportions remained high at 80\% and 66\%, respectively. 

\begin{tcolorbox}[
    colback=green!30!black!10,
    colframe=black!70,
    boxrule=0.8pt,
    arc=2mm,
    left=3mm,
    right=3mm,
    top=2mm,
    bottom=2mm,
    enhanced
]
\textbf{RQ1---Requirements Quality. }The stakeholder-centric RE practices were effective in producing a requirements
specification that stakeholders largely agreed to be understandable and correct. 
\end{tcolorbox}

\subsection{RQ2: Gap-analysis consistency}

RQ2 asked to what extent the resulting requirements specification enabled consistent stakeholder judgements about whether each requirement was already supported by the legacy system. This was evaluated through the additional questionnaire item on legacy-system coverage, again analysed through inter-rater agreement.

As shown in Table~\ref{tab:alpha-values}, agreement on functionality mapping was the lowest of all measured dimensions, with an alpha value of 0.45. This indicates only limited consistency in stakeholder judgements about whether requirements were already covered by the system-\textit{as-is}. 
The descriptive results in Table~\ref{tab:srs-values} reinforce this interpretation. Under the tight interpretation, no requirement achieved unanimous agreement regarding functionality mapping, and even under the lenient interpretation only 50\% of requirements were judged as mapped consistently enough to support a majority-based interpretation. 

The findings therefore suggest that the main limitation was not the understandability of the elicited requirements, but rather the difficulty of establishing a shared view of legacy system coverage. Stakeholders appeared to hold different understandings of what the existing system  actually supported. This is consistent with the well-known ambiguity and erosion of knowledge surrounding long-lived legacy systems~\cite{khadka2014professionals}, but also with the inherently partial viewpoints of different stakeholder roles~\cite{nuseibeh1996method}. Although not a novel finding per-se, it quantifies the \textit{magnitude} of the viewpoint differences.
The divergence in stakeholder viewpoints is also reflected in the prioritisation results, where agreement was reached for only 20\% of the requirements. However, this specific result was considered surprising, given that a Delphi method-based workshop was specifically organised after requirements elicitation interviews with the specific objective of achieving a shared understanding of the target system's goals. 

The results in Table~\ref{tab:srs-values} also highlight that the legacy system was generally not considered to include the majority of the envisioned functionality. On the one hand, this suggests that the renewal process was actually highly needed, as several needs were not currently satisfied. On the other hand, it might indicate that the way the RE practices were implemented tended to target \textit{novel} needs, leading stakeholders to focus on what is missing, rather than envisioning the complete system. 


\begin{tcolorbox}[
    colback=green!30!black!10,
    colframe=black!70,
    boxrule=0.8pt,
    arc=2mm,
    left=3mm,
    right=3mm,
    top=2mm,
    bottom=2mm,
    enhanced
]
\textbf{RQ2---Gap Analysis Consistency.} The resulting requirements specification enabled gap analysis in the sense that stakeholders could use it to assess legacy system coverage requirement by requirement. However, the relatively low agreement shows that this assessment was not sufficiently consistent to support strong conclusions about reuse potential. Furthemore, while the implementation of the RE approaches supported the identification of novel requirements, they were not effective in eliciting requirements already satisfied by the existing system.
\end{tcolorbox}

\section{Specifying Learning: Discussion}
\label{sec:discussion}


The following observations outline reflections that emerged during the final plenary workshop. The reflections were further elaborated by the authors based on their experience and the existing literature. At the end of each paragraph, we provide some actionable suggestions (using the symbol $\blacktriangleright$), to be assessed in future studies.  

\textit{Stakeholder Needs and the System-as-is.} 
Overall, the findings show that the intervention was successful in producing a requirements specification that stakeholders largely considered understandable and, to a considerable extent, correct (RQ1), but less successful in establishing agreement on  functionality mapping to the legacy systems (RQ2). This suggests that the main challenge was not only eliciting requirements for the system-\textit{to-be}, for which substantial disagreements were already observed during prioritisation, but also reconciling partially different stakeholder views on what the legacy systems already provided. From this perspective, the main value of the gap-analysis exercise was diagnostic: it revealed fragmented knowledge across stakeholders, and showed that legacy system modernisation requires, first and foremost, a reconstruction of a shared understanding of the existing systems. This apparently confirms the perspective of part of the literature~\cite{wolfart2021modernizing}, in which legacy system understanding is the first step of modernisation endeavours\footnote{Interestingly, Wolfart et al.~\cite{wolfart2021modernizing} derives a general process from the literature, and the elicitation of stakeholder needs is \textit{not} included in the process, remarking the scarce attention of the literature to the RE dimension.}. On the other hand, it should be noted that, in the case discussed, the initial uncertainty of the problem, and the similarity with the industrial experience from Khadka et al.~\cite{Khadka2013Migrating}, triggered the authors to start from stakeholders' needs, rather than the \textit{status quo}, i.e., the legacy system. The participants agreed that this was likely a mistake that should have been prevented, e.g., by dedicating more time during the interviews to understand the current system. However, RE practices appear to be strongly biased towards the development of novel systems, and even in the cases in which interview guidelines---also used in our interview design---recommend to understand the system/process-\textit{as-is}~\cite{donati2017common,ferrari2019learning}, this mostly serves the goal of identifying the \textit{problems} with the current system, rather than the features that actually work and should be retained. Novel guidelines are needed that specifically target the understanding and reconstruction of the system-\textit{as-is}, not only in terms of functionality, as done in the literature, but also in terms of needs that it already satisfies. Understanding the rationale underlying legacy system behaviour can help identify features that are no longer required, as the conditions or needs that originally motivated them may no longer apply. 

$\blacktriangleright$ Requirements elicitation in legacy system renewal should adopt a two step interview process with associated scripts, one to reconstruct the system-\textit{as-is} (including both strenghts and weaknesses), and the other one oriented to define novel requirements for the system-\textit{to-be}.

\textit{From Stakeholder-centric RE to Evidence-based RE.} When stakeholder perspectives diverge strongly and knowledge about legacy functionality has eroded over time, further progress appears to require complementary practices, both stakeholder-centric and system- and data-centric. 
These may include additional rounds of reconciliation among stakeholders, and, more importantly, systematic examination of evidence from the legacy systems themselves, such as existing reports, configurations, process traces, documentation, or implementation artefacts. These forms of triangulation can help move the renewal effort from a stakeholder-perceived view of the current and desired situations toward a more evidence-based and collectively validated understanding. While a stakeholder-centric RE process like the one enacted can bring the organisation to the point where major ambiguities become explicit,  additional evidence-oriented and consensus-building activities are needed to enable robust modernisation decisions.

$\blacktriangleright$ In legacy system modernisation, elicit information from stakeholders, but make sure to contrast opinions and viewpoints with system evidence, e.g., reports, documentation, logs, and artefacts. 

\textit{The Role of Automation and Large Language Models (LLMs).} To support evidence-based modernisation, automation is expected to play an important role, especially given the typical size and layered evolution of legacy systems, where functionality, design decisions, and operational assumptions accumulate over time, much like strata in an archaeological site. This suggests the need for approaches that integrate stakeholder-centric RE practices with automated system-understanding techniques. Relevant examples include \emph{feature location}, which aims to identify whether and where a given feature is implemented in a software system~\cite{dit2013feature}; \emph{specification mining}, which aims to infer likely behavioural rules or specifications from executions or software artefacts~\cite{kang2021adversarial}; \emph{process mining}, which reconstructs observed business or operational processes from event data and execution logs~\cite{van2022process}; and \emph{architecture recovery}, including mining software repositories for architectural knowledge, which aims to reconstruct higher-level structural views of a system from source code and related artefacts~\cite{soliman2025mining}. In addition, recent studies suggest that large language models (LLMs) may support automated architecture recovery~\cite{amalfitano2025automated,boronat2025mdre}, thereby offering a promising complement to more established reverse-engineering techniques. Together, these techniques point toward hybrid approaches in which stakeholder input is complemented with automatically recovered evidence about the system. 

$\blacktriangleright$ When analysing system evidence, exploit LLMs and automation in general, using recent research solutions for architecture recovery, specification mining, and process mining. 

$\blacktriangleright$ The RE community should engage more closely with the mining software repositories and software architecture communities to develop joint strategies for evidence-based legacy system modernisation.

\textit{Stakeholder Unavailability and Insufficient Evidence.}  In our study we experienced first-hand that stakeholder-centric RE is subject to inherent practical limitations, which may have strongly affected the results of the study. In industrial settings, including the one considered in this study, stakeholder availability is often limited, so participation across activities is necessarily based on sampling, representation, and best effort rather than complete and continuous involvement of all identified stakeholders. Consequently, even repeated activities with intense stakeholder involvement may fail to produce a fully shared understanding. This was visible in our case, where disagreement persisted despite several opportunities for interaction. We have suggested that evidence-based RE could address the problem. However, in the present case, evidence was only available to a limited extent, and the main repository of knowledge about the legacy systems were the stakeholders themselves. This suggests the need for RE practices that explicitly account for knowledge uncertainty and can support successful renewal despite limited stakeholder involvement and limited available evidence. 

$\blacktriangleright$ When the sources of evidence are limited, make sure to explicitly account for uncertainty in knowledge and requirements, e.g., associating  the different information items elicited with both stakeholder agreement and evidence support scores. 

\textit{The Need for Agile Renewal.} A main reflection of the action research team is that, to address the problem of stakeholder availability and limited evidence, stakeholder-centric RE for legacy system modernisation should be organised as an iterative and incremental learning process rather than as a one-shot planning exercise. Because stakeholder availability is inherently limited, involvement must often rely on representative participation, best effort, and progressive inclusion of those stakeholders who are most likely to reduce the most critical remaining uncertainties. In such a process, renewal would proceed stepwise, focusing on small slices of functionality and revising priorities as new understanding emerges. Where conflicting stakeholder accounts remain unresolved, the process should draw on additional evidence from the legacy systems themselves whenever such evidence is available. This suggests a direction that is compatible with agile and backlog-driven ways of working, for example by maintaining a living renewal backlog of requirements, ambiguities, evidence needs, and candidate migration slices. In this sense, approaches inspired by Scrum or other agile methods may be useful as templates for structuring renewal into short cycles of elicitation, validation, implementation planning, and reassessment.

$\blacktriangleright$ Agile software engineering techniques should be tailored to legacy system modernisation to ensure that project uncertainty is incrementally reduced. By introducing innovations through controlled iterations, these tailored practices yield tangible product increments that directly facilitate stakeholder consensus.

$\blacktriangleright$ Modernisation-specific backlogs should be defined that account for existing, desired, and obsolete features, with associated uncertainty measures or qualifiers. 

\section{Threats to Validity}
\label{sec:threats}

\paragraph{Internal Validity}
The study was conducted as an action research intervention in a real industrial setting, which makes it difficult to isolate the effect of each individual RE practice on the final outcome. The observed results stem from the combination of stakeholder identification, interviews, thematic analysis, prioritisation, and questionnaire-based evaluation, rather than from any single step in isolation. Moreover, the order of the activities, although justified by previous experiences~\cite{Khadka2013Migrating} and by established RE literature~\cite{pohl2025requirements}, could have affected the results. Stakeholder participation also varied across activities, and not all identified stakeholders took part in all phases. As a consequence, the resulting requirements specification and evaluation outcomes may have been influenced by who was available, how actively they participated, and how representative their contributions were of the broader organisation.

\paragraph{External Validity}
The study concerns a single modernisation effort in one multinational energy company, focused on a specific reporting-related legacy system. The organisational setting, the nature of the system, the available stakeholders, and the company’s ongoing strategy all shaped the intervention. For this reason, the findings cannot be assumed to generalise directly to other organisations or domains. In particular, the suitability of the selected RE practices may vary depending on system complexity, the availability of documentation and other evidence, the maturity of the organisation’s processes, and the extent to which legacy knowledge is still held by stakeholders. The study should therefore be interpreted as an in-depth industrial investigation that provides analytical rather than statistical generalisation. Nevertheless, according to case-based generalisation principles~\cite{wieringa2015six}, our conclusions might be applicable to contexts that share similar characteristics with our study settings: multi-national energy companies dealing with system modernisation; large legacy systems for data analytics and reporting; involvement of diverse stakeholders, at management, user, and IT level. 

\paragraph{Construct Validity}
Several constructs in the study were operationalised in ways that may only partially capture the phenomena of interest. First, the effectiveness of the intervention was assessed through two main lenses: the perceived quality of the resulting requirements specification and its usefulness for supporting requirements-based gap analysis. While these are reasonable operationalisations for this study, they do not exhaust all possible notions of effectiveness in legacy system modernisation. Second, requirement quality was assessed through understandability and correctness, which are only a limited set of the quality attributes outlined by Davis et al.~\cite{davis2006effectiveness}. Third, understandability was merely estimated by asking users whether they considered the requirements understandable. While this self-assessment may not fully capture underlying comprehension barriers in a multi-lingual environment, the risk of misinterpretation was mitigated because the requirements were written using a simple language.  Fourth, gap analysis was operationalised through stakeholder judgements of whether each elicited requirement was already fully, partially, or not supported by the legacy system. This captures one important aspect of gap analysis, but not broader process, organisational, or architectural gaps. 

The qualitative parts of the study introduce further construct-related uncertainty, since coding and thematic analysis necessarily involve researcher interpretation. The Delphi workshop mitigated this threat to some extent, but could not eliminate it entirely. Finally, the agreement analysis should be interpreted with caution, as the chosen operationalisation of inter-rater agreement reflects observed consistency in this setting rather than a fully theory-neutral measure of underlying truth.
\section{Conclusion}
\label{sec:conclusion}

This paper investigated the role of stakeholder-centric RE practices in legacy system modernisation. Through an action research study in a multinational energy company, we found that these practices were paramount in producing requirements that stakeholders largely considered understandable and correct. 
However, when the resulting requirements were used to assess which ones were already supported by the legacy system, substantial disagreement emerged, pointing to fragmented understanding of the existing system. This suggests both that stakeholder-centric practices, as implemented in our study, were insufficient on their own, and that elicitation requires stronger support from automated techniques combined with human judgement to understand the system-\textit{as-is}.
The need for a stronger human + automated RE combo is bound to be required for large modernisation where employee turnover and any other organisational rewiring exercise change the project knowledge structure to a point where drift between old and new becomes unmanageable. This would also lead to assume that an organisation/community-centric approach---as opposed to a stakeholder-centric one---may prove beneficial in the future, as systems become larger and more unmanageable.

The industrial participants agreed that stakeholder-centric RE practices are a non-negotiable starting point. While they were not sufficient on their own to resolve conflicting interpretations of  the actual behaviour of the legacy systems, they gave solid grounds to progress in the modernisation efforts. They allowed them to make their idea more explicit and gave them the opportunity to see that an agreement and knowledge problem existed. 
Requirements elicitation research should pose a stronger emphasis on the pre-existing system as a source of requirements, and RE techniques should be developed to elicit and combine stakeholder and system knowledge, possibly captured through automated methods. The goal of RE for legacy system is not to fully resolve the renewal problem in one pass, but to progressively reduce uncertainty and build a sufficiently evidence-based basis for modernisation decisions.

\smallskip
\noindent
\textbf{AI Disclosure.} ChatGPT 5.4 was used to assist in revising the language and presentation of the manuscript. The authors take full responsibility for the final text.

%
\IEEEpeerreviewmaketitle

\bibliographystyle{IEEEtran}
\bibliography{biblio}

@article{wieringa2015six,
  title={Six strategies for generalizing software engineering theories},
  author={Wieringa, Roel and Daneva, Maya},
  journal={Science of computer programming},
  volume={101},
  pages={136--152},
  year={2015},
  publisher={Elsevier}
}

@inproceedings{Alexandrova2015The,
  author       = {Alexandrova, A. and Rapanotti, L. and Horrocks, I.},
  title        = {The Legacy Problem in Government Agencies: An Exploratory Study},
  booktitle    = {dg.o'15},
  year         = {2015},
  pages        = {150--159},
  doi          = {10.1145/2757401.2757406}
}

@techreport{Bisbal1997A,
  author      = {Bisbal, J. and Lawless, D. and Wu, B. and Grimson, J. and Wade, V. and Richardson, R. and O'Sullivan, D.},
  title       = {A Survey of Research into Legacy System Migration},
  institution = {Trinity College Dublin},
  year        = {1997}
}

@inproceedings{Hasselbring2004The,
  author    = {Hasselbring, W. and Reussner, R. and Jaekel, H. and Schlegelmilch, J. and Teschke, T. and Krieghoff, S.},
  title     = {The Dublo Architecture Pattern for Smooth Migration of Business Information Systems: An Experience Report},
  booktitle = {ICSE'04},
  year      = {2004},
  pages     = {117--126},
  doi       = {10.1109/ICSE.2004.1317434}
}

@techreport{Visaggio1997Comprehending,
  author      = {Visaggio, G.},
  title       = {Comprehending the Knowledge Stored in Aged Legacy Systems to Improve Their Qualities with a Renewal Process},
  institution = {International Software Engineering Research Network},
  number      = {97-26},
  year        = {1997}
}

@book{staron2020action,
  title={Action research in software engineering},
  author={Staron, Miroslaw},
  year={2020},
  publisher={Springer}
}

@article{Hasan2023Legacy,
  title        = {Legacy systems to cloud migration: A review from the architectural perspective},
  author       = {Hasan, Muhammad Hafiz and Osman, Mohd Hafeez and Admodisastro, Novia Indriaty and Muhammad, Muhamad Sufri},
  journal      = {JSS},
  volume       = {202},
  year         = {2023},
  month        = aug,
  pages        = {111702},
  doi          = {10.1016/j.jss.2023.111702}
}

@inproceedings{fritzsch2018monolith,
  title={From monolith to microservices: A classification of refactoring approaches},
  author={Fritzsch, Jonas and Bogner, Justus and Zimmermann, Alfred and Wagner, Stefan},
  booktitle={DEVOPS'18},
  pages={128--141},
  year={2018},
  organization={Springer}
}

@inproceedings{wolfart2021modernizing,
  author    = {Daniele Wolfart and Wesley K. G. Assun{\c{c}}{\~a}o and Ivonei F. da Silva and Diogo C. P. Domingos and Ederson Schmeing and Guilherme L. Donin Villaca and Diogo do Nascimento Paza},
  title     = {Modernizing Legacy Systems with Microservices: A Roadmap},
  booktitle = {EASE'21},
  pages     = {149--159},
  year      = {2021},
  publisher = {ACM},
  doi       = {10.1145/3463274.3463334}
}

@inproceedings{olanrewaju2024investigating,
  title={Investigating systems modernisation: approaches, challenges and risks},
  author={Hogan, Gareth and Shalkauskaite, Patricija and Zhu, Mengte and Derwin, Martin and Yilmaz, Murat and McCarren, Andrew and Clarke, Paul M},
  booktitle={EuroSPI'24},
  pages={147--162},
  year={2024},
  organization={Springer}
}

@article{assunccao2025contemporary,
  title={Contemporary software modernization: Strategies, driving forces, and research opportunities},
  author={Assun{\c{c}}{\~a}o, Wesley KG and Marchezan, Luciano and Arkoh, Lawrence and Egyed, Alexander and Ramler, Rudolf},
  journal={TOSEM},
  volume={34},
  number={5},
  pages={1--35},
  year={2025},
  publisher={ACM New York, NY}
}

@inproceedings{davis2006effectiveness,
  title={Effectiveness of requirements elicitation techniques: Empirical results derived from a systematic review},
  author={Davis, Alan and Dieste, Oscar and Hickey, Ann and Juristo, Natalia and Moreno, Ana M},
  booktitle={RE'06},
  pages={179--188},
  year={2006},
  organization={IEEE}
}

@inproceedings{khadka2014professionals,
  title={How do professionals perceive legacy systems and software modernization?},
  author={Khadka, Ravi and Batlajery, Belfrit V and Saeidi, Amir M and Jansen, Slinger and Hage, Jurriaan},
  booktitle={ICSE'14},
  pages={36--47},
  year={2014}
}

@article{sneed2019reimplementing,
  title={Re-implementing a legacy system},
  author={Sneed, Harry and Verhoef, Chris},
  journal={JSS},
  volume={155},
  pages={162--184},
  year={2019},
  publisher={Elsevier}
}

@book{pohl2025requirements,
  author    = {Klaus Pohl},
  title     = {Requirements Engineering},
  subtitle  = {Fundamentals, Principles, and Techniques},
  edition   = {2},
  publisher = {Springer},
  address   = {Berlin, Heidelberg},
  year      = {2025},
  isbn      = {978-3-662-69204-2},
  pagetotal = {889},
  note      = {XVIII, 889 pages},
}

@article{stol2018abc,
  title={The ABC of software engineering research},
  author={Stol, Klaas-Jan and Fitzgerald, Brian},
  journal={TOSEM},
  volume={27},
  number={3},
  pages={1--51},
  year={2018},
  publisher={ACM New York, NY, USA}
}

@article{martinez2017managing,
  title={Managing legacy system costs: A case study of a meta-assessment model to identify solutions in a large financial services company},
  author={Crotty, James and Horrocks, Ivan},
  journal={API},
  volume={13},
  number={2},
  pages={175--183},
  year={2017},
  publisher={Elsevier}
}

@inproceedings{garcia2023lessons,
  title={Lessons Learned in Model-Based Reverse Engineering of Large Legacy Systems},
  author={Garc{\'\i}a-Borgo{\~n}{\'o}n, Laura and Barcelona, Miguel Angel and Egea, Armando J and Reyes, German and Sainz-de-la-maza, Alejandro and Gonz{\'a}lez-Uzabal, Adolfo},
  booktitle={CAiSE},
  pages={330--344},
  year={2023},
  organization={Springer}
}

@article{mohottige2025reengineering,
  author    = {Thakshila Imiya Mohottige and Artem Polyvyanyy and Colin J. Fidge and Rajkumar Buyya and Alistair Barros},
  title     = {Reengineering software systems into microservices: State-of-the-art and future directions},
  journal   = {IST},
  volume    = {183},
  pages     = {107732},
  year      = {2025},
  publisher = {Elsevier}
}

@incollection{carvalho2023re,
  author    = {Luiz Carvalho and Alessandro F. Garcia and Wesley K. G. Assun{\c{c}}{\~a}o and Thelma Elita Colanzi and Rodrigo Bonif{\'a}cio and Leonardo P. Tizzei and Rafael Maiani de Mello and Renato Cerqueira and M{\'a}rcio Ribeiro and Carlos Lucena},
  title     = {Re-engineering Legacy Systems as Microservices: An Industrial Survey of Criteria to Deal with Modularity and Variability of Features},
  booktitle = {Handbook of Re-Engineering Software Intensive Systems into Software Product Lines},
  pages     = {471--494},
  year      = {2023},
  publisher = {Springer},
  doi       = {10.1007/978-3-031-11686-5_19}
}

@book{seacord2003modernizing,
  title={Modernizing legacy systems: software technologies, engineering processes, and business practices},
  author={Seacord, Robert C and Plakosh, Daniel and Lewis, Grace A},
  year={2003},
  publisher={Addison-Wesley Professional}
}

@inproceedings{marquez2015framework,
  title={A framework for secure migration processes of legacy systems to the cloud},
  author={M{\'a}rquez, Luis and Rosado, David G and Mouratidis, Haralambos and Mellado, Daniel and Fern{\'a}ndez-Medina, Eduardo},
  booktitle={CAiSE},
  pages={507--517},
  year={2015},
  organization={Springer}
}

@inproceedings{grieger2016concept,
  title={Concept-based engineering of situation-specific migration methods},
  author={Grieger, Marvin and Fazal-Baqaie, Masud and Engels, Gregor and Klenke, Markus},
  booktitle={International Conference on Software Reuse},
  pages={199--214},
  year={2016},
  organization={Springer}
}

@article{dit2013feature,
  title={Feature location in source code: a taxonomy and survey},
  author={Dit, Bogdan and Revelle, Meghan and Gethers, Malcom and Poshyvanyk, Denys},
  journal={Journal of software: Evolution and Process},
  volume={25},
  number={1},
  pages={53--95},
  year={2013},
  publisher={Wiley Online Library}
}

@article{kang2021adversarial,
  title={Adversarial specification mining},
  author={Kang, Hong Jin and Lo, David},
  journal={TOSEM},
  volume={30},
  number={2},
  pages={1--40},
  year={2021},
  publisher={ACM New York, NY, USA}
}

@book{van2022process,
  title={Process mining handbook},
  author={van der Aalst, Wil MP and Carmona, Josep},
  volume={448},
  year={2022},
  publisher={Springer}
}

@article{nuseibeh1996method,
  title={Method engineering for multi-perspective software development},
  author={Nuseibeh, Bashar and Finkelstein, Anthony and Kramer, Jeff},
  journal={IST},
  volume={38},
  number={4},
  pages={267--274},
  year={1996},
  publisher={Elsevier}
}

@inproceedings{boronat2025mdre,
  title={MDRE-LLM: A tool for analyzing and applying llms in software reverse engineering},
  author={Boronat, Artur and Mustafa, Jawad},
  booktitle={SANER'25},
  pages={850--854},
  year={2025},
  organization={IEEE}
}

@inproceedings{amalfitano2025automated,
  title={Automated Software Architecture Design Recovery from Source Code Using LLMs},
  author={Amalfitano, Domenico and De Luca, Marco and Santilli, Tiziano and Pelliccione, Patrizio and Fasolino, Anna Rita},
  booktitle={ECSA'25},
  pages={73--89},
  year={2025},
  organization={Springer}
}

@article{soliman2025mining,
  title={Mining software repositories for software architecture—A systematic mapping study},
  author={Soliman, Mohamed and Albonico, Michel and Malavolta, Ivano and Wortmann, Andreas},
  journal={IST},
  volume={181},
  pages={107677},
  year={2025},
  publisher={Elsevier}
}

@inproceedings{NuseibehEasterbrook2000,
  author    = {Bashar Nuseibeh and Steve Easterbrook},
  title     = {Requirements Engineering: A Roadmap},
  booktitle = {FOSE'00},
  year      = {2000},
  pages     = {35--46},
  doi       = {10.1145/336512.336523}
}

@article{Palomares2021Elicitation,
  title   = {The state-of-practice in requirements elicitation: an extended interview study at 12 companies},
  author  = {Palomares, Cristina and Franch, Xavier and Quer, Carme and Chatzipetrou, Panagiota and L{\'o}pez, Lidia and Gorschek, Tony},
  journal = {REJ},
  volume  = {26},
  pages   = {273--299},
  year    = {2021},
  doi     = {10.1007/s00766-020-00345-x}
}

@article{Visaggio2001Ageing,
  author    = {Visaggio, G.},
  title     = {Ageing of a Data-Intensive Legacy System: Symptoms and Remedies},
  journal   = {J. Softw. Maint.: Res. Pract.},
  volume    = {13},
  year      = {2001},
  pages     = {281--308},
  doi       = {10.1002/smr.234}
}

@inproceedings{Battaglia1998Renaissance,
  author    = {Battaglia, M. and Savoia, G. and Favaro, J.},
  title     = {Renaissance: A Method to Migrate from Legacy to Immortal Software Systems},
  booktitle = {CSMR'98},
  year      = {1998},
  pages     = {197--200},
  doi       = {10.1109/CSMR.1998.665807}
}

@inproceedings{Khadka2013Migrating,
  author    = {Khadka, R. and Saeidi, A. and Jansen, S. and Hage, J. and Haas, G. P.},
  title     = {Migrating a Large Scale Legacy Application to SOA: Challenges and Lessons Learned},
  booktitle = {20th Working Conference on Reverse Engineering},
  year      = {2013},
  pages     = {425--432},
  doi       = {10.1109/WCRE.2013.6671318}
}

@article{Friedman2002Developing,
  author    = {Friedman, A. L. and Miles, S.},
  title     = {Developing Stakeholder Theory},
  journal   = {Journal of Management Studies},
  volume    = {39},
  number    = {1},
  year      = {2002},
  pages     = {1--21}
}

@article{Pacheco2012A,
  author    = {Pacheco, C. and Garcia, I.},
  title     = {A Systematic Literature Review of Stakeholder Identification Methods in Requirements Elicitation},
  journal   = {JSS},
  volume    = {85},
  number    = {9},
  year      = {2012},
  pages     = {2171--2181},
  doi       = {10.1016/j.jss.2012.04.075}
}

@article{Hujainah2018Software,
  author    = {Hujainah, F. and Bakar, R. B. A. and Abdulgabber, M. A. and Zamli, K. Z.},
  title     = {Software Requirements Prioritization: A Systematic Literature Review on Significance, Stakeholders, Techniques and Challenges},
  journal   = {IEEE Access},
  volume    = {6},
  year      = {2018},
  pages     = {71497--71523},
  doi       = {10.1109/ACCESS.2018.2881755}
}

@article{ferrari2016ambiguity,
  title={Ambiguity and tacit knowledge in requirements elicitation interviews},
  author={Ferrari, Alessio and Spoletini, Paola and Gnesi, Stefania},
  journal={REJ},
  volume={21},
  number={3},
  pages={333--355},
  year={2016},
  publisher={Springer}
}

@inproceedings{ferrari2019learning,
  title={Learning requirements elicitation interviews with role-playing, self-assessment and peer-review},
  author={Ferrari, Alessio and Spoletini, Paola and Bano, Muneera and Zowghi, Didar},
  booktitle={2019 IEEE 27th international requirements engineering conference (RE)},
  pages={28--39},
  year={2019},
  organization={IEEE}
}

@inproceedings{donati2017common,
  title={Common mistakes of student analysts in requirements elicitation interviews},
  author={Donati, Beatrice and Ferrari, Alessio and Spoletini, Paola and Gnesi, Stefania},
  booktitle={REFSQ'17},
  pages={148--164},
  year={2017},
  organization={Springer}
}

@inproceedings{Davis1993Identifying,
  author    = {Davis, A. and Overmyer, S. and Jordan, K. and Caruso, J. and Dandashi, F. and Dinh, A. and Kincaid, G. and Ledeboer, G. and Reynolds, P. and Sitaram, P. and Ta, A. and Theofanos, M.},
  title     = {Identifying and Measuring Quality in a Software Requirements Specification},
  booktitle = {METRICS'93},
  year      = {1993},
  pages     = {141--152},
  doi       = {10.1109/METRIC.1993.263792}
}

@misc{Krippendorff2011Computing,
  author    = {Krippendorff, K.},
  title     = {Computing Krippendorff's Alpha-Reliability},
  year      = {2011},
  howpublished = {Departmental Papers (ASC)},
  url       = {https://repository.upenn.edu/asc_papers/43}
}

@article{Alshenqeeti2014Interviewing,
  author    = {Alshenqeeti, H.},
  title     = {Interviewing as a Data Collection Method: A Critical Review},
  journal   = {English Linguistics Research},
  volume    = {3},
  number    = {1},
  year      = {2014},
  pages     = {39--45},
  doi       = {10.5430/elr.v3n1p39}
}

@incollection{Barbour2005Interviewing,
  author    = {Barbour, R. and Schostak, J. F.},
  title     = {Interviewing and Focus Groups},
  booktitle = {Research Methods in the Social Sciences},
  publisher = {Sage Publications},
  year      = {2005},
  pages     = {41--48}
}

@incollection{Braun2012Thematic,
  author    = {Braun, V. and Clarke, V.},
  title     = {Thematic Analysis},
  booktitle = {APA Handbook of Research Methods in Psychology},
  editor    = {Cooper, H. and Camic, P. M. and Long, D. L. and Panter, A. T. and Rindskopf, D. and Sher, K. J.},
  volume    = {2},
  publisher = {American Psychological Association},
  year      = {2012},
  pages     = {57--71}
}

@article{Moogk2012Minimum,
  author    = {Moogk, D. R.},
  title     = {Minimum Viable Product and the Importance of Experimentation in Technology Startups},
  journal   = {Technology Innovation Management Review},
  year      = {2012},
  pages     = {23--26}
}

@inproceedings{Schuh2018Agile,
  author    = {Schuh, G. and Doelle, C. and Schloesser, S.},
  title     = {Agile Prototyping for Technical Systems: Towards an Adoption of the Minimum Viable Product Principle},
  booktitle = {Proceedings of NordDesign 2018},
  year      = {2018}
}

@article{Suchetha2024, 
title={Assessing the Effectiveness of MoSCoW Prioritization in Software Development: A Holistic Analysis across Methodologies}, 
volume={10}, 
DOI={10.4108/eetiot.6515}, 
journal={TIoT}, 
author={Suchetha Vijayakumar and K. Krishna-Prasad and Holla M., Raviraja}, 
year={2024}, 
month={Oct.} 
}
\end{document}